\documentclass[twocolumn,aps,pre,showpacs]{revtex4}
\usepackage[latin1]{inputenc}
\usepackage{graphicx}
\usepackage{latexsym}
\usepackage{amssymb}
\usepackage{amsmath}

\def\d{{\rm d}}

\def\w{{\bf w}}
\def\x{{\bf x}}

\def\z{{\bf z}}

\def\beq{\begin{eqnarray}}
\def\eeq{\end{eqnarray}}
\begin{document}

\title{Forcing anomalous scaling on demographic fluctuations}
\author{Piero Olla}
\affiliation{ISAC-CNR and INFN, Sez. Cagliari, I--09042 Monserrato, Italy.}
\date{\today}

\begin{abstract}
We discuss the conditions under which a population of anomalously diffusing 
individuals can be characterized by demographic fluctuations that are
anomalously scaling themselves. Two examples are provided in the case
of individuals migrating by Gaussian diffusion, and by a 
sequence of L\'evy flights.
\end{abstract}

\maketitle
\section{Introduction}
Simple birth-death models, with individual migration described
by a diffusion process, provide a simple illustration of how demographic
stochasticity at the microscopic level translates into the phenomenon
of spatial clustering \cite{zhang90,meyer96}.
An initial uniform distribution of individuals will develop spatial fluctuations
whose amplitude grows in time without bound. This phenomenon requires 
the individual birth and death rates to be 
equal and constant (neutral conditions), and that there are no interactions
among individuals. These conditions may look somewhat artificial. Nonetheless,
situations of this sort have been realized in the laboratory, and have allowed
experimental measurement of clustering in living populations \cite{houchmandzadeh08}.

The original setting of model, such as those considered in  \cite{zhang90,meyer96},
involved Brownian walkers (``Brownian bugs'', in the terminology
of \cite{young01}), but the generalization to the case
of anomalously diffusing individuals was considered as well \cite{heinsalu10,olla12}.
Contrary to what could have been expected, clustering
did not appear to be sensitive to the character of the
diffusion process. 
What appeared to be effective in modifying the population dynamics, instead, was
the possibility that memory of the trajectories be transferred from individuals to their
descendants \cite{olla12}. 

At least in the case of a migration dynamics of
the continuous time random walk (CTRW) type \cite{klafter87,montroll65}, 
memory transfer between generations
was insufficient to produce any scaling dependence
on the Hurst exponent of the diffusion in the population dynamics
\cite{olla12}.
What was observed, was collapse
on a Galton-Watson like behavior \cite{harris63}.
The question remained open on whether different
models would lead to the same result.

What we intend to analyze in the present report is precisely
under what conditions a dependence
between the  scaling of the local population fluctuations, and that of the diffusion process,
can exist. The analysis will be limited to a one-dimensional contest, but arguments
will be provided to generalize the results to $D>1$.
We shall consider two cases: that of anomalous diffusion produced by a Gaussian 
process, and that of diffusion produced by a sequence of L\'evy flights 
\cite{montroll79,bouchaud90}. 
These are situation of importance both conceptually and
from the point of view of applications.
Physically relevant examples of anomalous Gaussian diffusion include 
particles migrating with a
velocity that is solution of a generalized Langevin equation (GLE) \cite{zwanzig},
and, of course, the fractional Brownian motion (FBM) \cite{mandelbrot68}.
L\'evy flights constitute an important ingredient in modelling the
spreading of epidemies \cite{janssen99}, the spreading of seeds in a natural environment
\cite{clark99}, biologic searching strategies \cite{sims08},
as well as bacteria dynamics \cite{levandowsky97}. 

In the case of Gaussian diffusion, we shall see that ``anomalous'' scaling of the population 
fluctuation can indeed be produced, and it is the result of a delicate interplay between 
Gaussian statistics, memory transfer between generations, and the non-Markovian 
nature of the process. In the case of L\'evy flights, 
the Markovian nature of the
diffusion process makes any memory transfer between generations irrelevant
to the dynamics, and the scaling of the local population flucttuations will appear to be
determined automatically by the stable law of the L\'evy process.

\section{Gaussian diffusion}
Let us consider the simple situation of a population of non-interacting individuals, 
with identical birth and death rates $\Gamma_B=\Gamma_D\equiv\Gamma$
(one offspring for each birth event)
to insure stationarity. We assume births and
deaths to be Markovian and focus on the effect that anomalous diffusion of the
individuals may have at the population level. Let us limit the analysis,
for the moment, to a one-dimensional situation.

We consider first the case of a Gaussian, anomalous diffusion process. Contrary to the
case of normal diffusion, in which the condition of linear scaling of the mean square
displacement (assuming Gaussian statistics) determines uniquely the process, full 
knowledge of the displacement correlation is necessary in the anomalous 
case. 

Let us consider a population that initially is uniformly distributed in space, with 
$\rho_1(x,0)=n_0$.
Equality of the birth and death rates guarantees that
the population remains on the average uniformly distributed, with the same
population density $\rho_1(x,t)\equiv\langle n(x,t)\rangle=n_0$ [we
have indicated with $n(x,t)$ the instantaneous value of the population 
density, including fluctuations]. To study fluctuations, we must consider the
higher moments of $n(x,t)$, in particular the two-bug correlation
\beq
\rho_2(x_1,x_2;t)=\langle n(x_1,t) n(x_2,t)\rangle,
\label{eq1}
\eeq
or, more precisely, its
connected part $\rho_{2c}(x_1,x_2;t)=\rho_2(x_1,x_2;t)-n_0^2$.

\begin{figure}
\begin{center}
\includegraphics[draft=false,width=6.5cm]{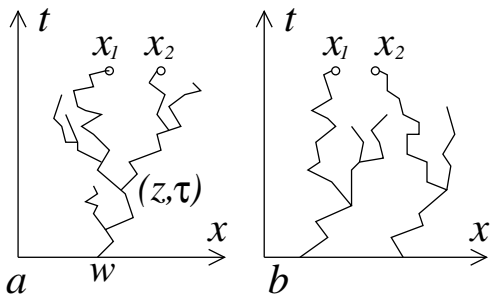}
\caption{Family tree contributing to the connected ($a$) and disconnected ($b$) 
part of the two-bug correlation $\rho_2(x_1,x_2;t)$.  [see Eq. (\ref{eq1})]. 
The correlation is obtained
counting pairs of branches passing through the sampling points at $x_{1,2}$.
}
\label{fbmfig1}
\end{center}
\end{figure}
A representation of the population dynamics is obtained considering the family trees
of its members, as illustrated in Fig.  \ref{fbmfig1}. 
A quantity such as the correlation $\rho_2(x_1,x_2;t)$ is obtained therefore by summing
over the tree branches passing through $(x_1,t)$ and $(x_2,t)$ and averaging over
the family tree configurations.
The family trees responsible for the connected component $\rho_{2c}(x_1,x_2;t)$ are
those represented in case $a$ of Fig. \ref{fbmfig1}, in which the individuals at
$(x_{1,2},t)$ share a common ancestor. Clearly, 
cutting the family tree of case $a$ above the branching at
$(z,\tau)$ and neglecting all the previous history,
would lead to a disconnected tree identical to the one represented in $b$.
It is easy to be convinced that
the same result would be obtained by reinitializing from scratch
the migration strategies of the individuals at the branching.
%
A complete reinitialization of migration
at each birth event, would lead  in fact to Markovianization of the population dynamics
at times longer than the individual lifetime $\Gamma^{-1}$, with the displacements
of different individuals over a lifetime becoming independent.
(In the case of migration by a CTRW,
this lead to a clustering
dynamics in the population identical to that of Brownian bugs \cite{houch02}).

In order to obtain a non-trivial dynamics, 
the new-born individuals must preserve some memory of the trajectories
followed by their parents.
We shall consider the situation in which individuals along the same family line
have full memory of the trajectory followed by their ancestors.

Let us
indicate by $G(x-z,t-\tau)\equiv\rho_1(x,t|z,\tau)$
the mean population density at $(x,t)$ given
presence of an ancestor at $(z,\tau)$, with $\tau<t$, and with 
$\rho_1(x,t|z,\tau;w,0)$, the same quantity conditioned to presence of a 
common ancestor at $(w,0)$. From equality of the birth and death rates, the 
total number of individuals accounted for by $\rho_1(x,t|.)$  is one and these
functions are normalized to one and coincide with the conditional
PDF's (probability density functions) for the position a single walker in 
the absence of birth and death effects.  We can write:
\beq
\rho_{2c}(x_1,x_2;t)&=&2\Gamma\int_0^t\d\tau\int\d z\int\d w\ 
\rho_1(x_1,t|z,\tau;w,0)
\nonumber
\\
&\times& \rho_1(x_2,t|z,\tau;w,0)G(z-w,\tau)
\nonumber
\\
&\times&\rho_1(w,0).
\label{eq2}
\eeq
We can read Eq. (\ref{eq2}) from the family tree $a$ in Fig. \ref{fbmfig1}:
\begin{itemize}
\item
The factor $G(z-w,\tau)\rho_1(w,0)$ is the individual density
at $(z,\tau)$, originating from an ancestor at $(w,0)$, that is available
to generate offspring; $\Gamma G(z-w,\tau)\rho_1(w,0)$ is the 
corresponding total rate of birth.
\item
The factors $\rho_1(x_{1,2},t|z,\tau;w,0)$ enforce memory of the position of
the common ancestor at $(w,0)$ on the offspring at $(z,\tau)$, until it
(or one of its descendants) reaches $x_{1,2}$. 
\item
The factor 2 on the right side accounts for the number of ways in which 
parent and offspring at $(z,\tau)$ (or their descendants) may distribute
between $x_{1,2}$. 
\end{itemize}
Gaussian statistics makes determination of the conditional PDF's
$\rho_1(x,t|.)$ straightforward \cite{note}:
\beq
G(z-w,\tau)=\frac{1}{(2\pi)^{1/2}\sigma(\tau)}
\exp\Big(-\frac{|z-w|^2}{2\sigma^2(\tau)}\Big)
\label{eq3}
\eeq
and
\beq
\rho_1(x,t|z,\tau;w,0)=\frac{1}{(2\pi)^{1/2}\sigma(t|\tau;z-w)}
\nonumber
\\
\times
\exp\Big(-\frac{|x-\mu(t|\tau,z-w)|^2}{2\sigma^2(t|\tau;z-w)}\Big),
\label{eq4}
\eeq
where the conditional mean and mean square displacements
$\mu(t|.)$ and $\sigma^2(t|.)$ are given by
\beq
\mu(t|\tau,z-w)=
w+\frac{\langle x(t)x(\tau)\rangle}{\langle x^2(\tau)\rangle}(z-w)
\label{eq5}
\eeq
and
\beq
\sigma^2(t|\tau;z-w)\equiv\sigma^2(t;\tau)=\sigma^2(t)-
\frac{\langle y(t)y(\tau)\rangle^2}{\sigma^2(\tau)},
\label{eq6}
\eeq
with $y(t)=x(t)-x(0)$ and $\sigma^2(t)=\langle y^2(t)\rangle$.

Substituting Eqs. (\ref{eq3}-\ref{eq6}) into Eq. (\ref{eq2}), and using
$\rho_1(x,0)=n_0$, we obtain the  result
\beq
\rho_{2c}(x_1,x_2;t)=\frac{\Gamma n_0}{2\sqrt{\pi}}\int_0^t\frac{\d\tau}{\sigma(t;\tau)}
\exp\Big[-\frac{(x_1-x_2)^2}{4\sigma^2(t,\tau)}\Big].
\label{eq7}
\eeq
For $t\gg\tau$ we see that $\sigma(t;\tau)\to\sigma(t)$.
The Markovian case is obtained substituting $\sigma(t;\tau)\to\sigma(t-\tau)$,
for $t,\tau$ generic,
in Eq. (\ref{eq7}), that becomes in this way the solution of a heat equation with 
constant Dirac delta forcing at $x_1=x_2$ \cite{houch02}. 

\begin{figure}
\begin{center}
\includegraphics[draft=false,width=6.5cm]{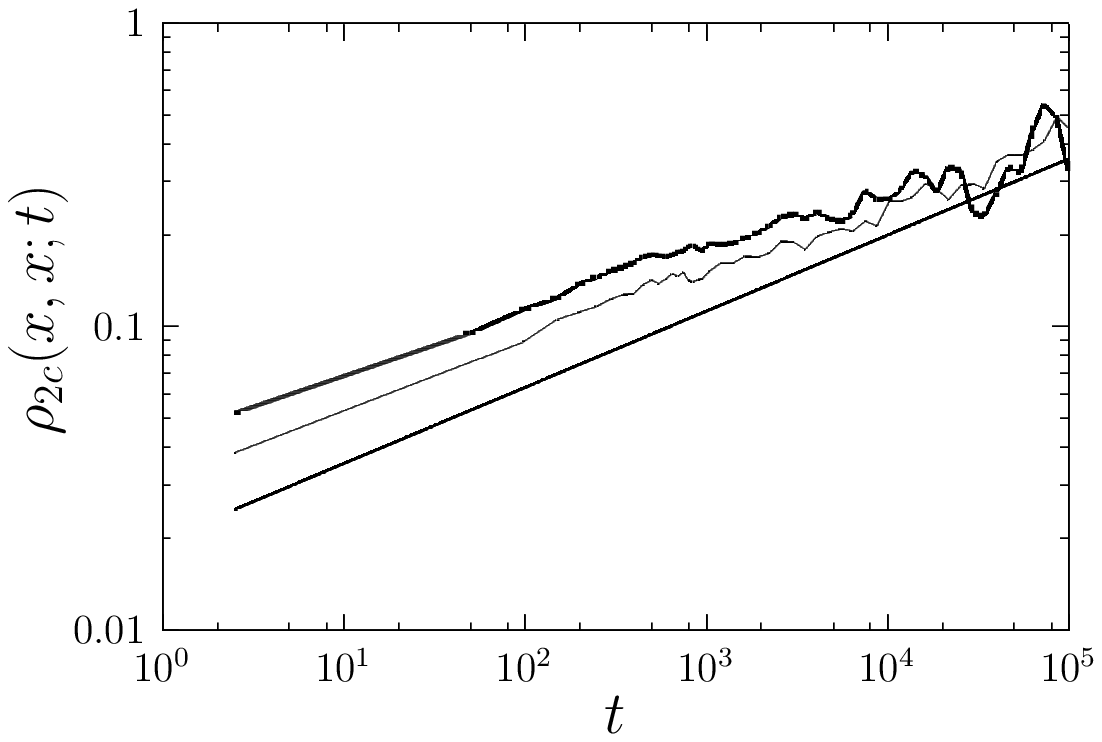}
\caption{
Fluctuation build-up in the case the individuals migrate by Gaussian diffusion 
(heavy line) and by L\'evy flights (thin line). The slope $t^{1-H}$, for $H=1/\beta=0.75$,
is shown for reference. In both cases $N=10^5$ individuals in a periodic domain 
were considered. In the time units considered: the fastest mode in the FBM has relaxation time
one; the discretization of the L\'evy flights is $\Delta t=0.25$; $\Gamma=0.1$.
}
\label{fbmfig2}
\end{center}
\end{figure}

The scaling of the square amplitude fluctuation $\rho_{2c}(x,x;t)$ can be calculated
explicitly if the correlation profile entering Eq. (\ref{eq6}) is known. In the 
case of an FBM \cite{mandelbrot68}:
\beq
\langle y(t)y(\tau)\rangle=(\kappa_H/2)(t^{2H}+\tau^{2H}-|t-\tau|^{2H})
\label{eq8}
\eeq
and we get
\beq
\rho_{2c}(x,x;t)=C\Gamma n_0\kappa_H^{1/2}t^{1-H},
\label{eq9}
\eeq
with
\beq
C=\frac{1}{2\sqrt{\pi}}\int_0^1\frac{u^H\d u}{\sqrt{u^{2H}-[1+u^{2H}-(1-u)^{2H}]^2/2}}.
\nonumber
\eeq
The general condition $\sigma(t;\tau)\to\sigma(t)$ for $t/\tau$ large,
leads us to expect that the same scaling of Eq. (\ref{eq9}) be recovered also
for other Gaussian, anomalous diffusion processes \cite{nota0}. 
An example is illustrated
in Fig. \ref{fbmfig2}, in the case of a superdiffusive process generated with 
the algorithm described in \cite{biferale97}, in which the velocity
of the individuals is obtained as a superposition of Ornstein-Uhlenbeck processes.

From Eq. (\ref{eq7}) we could also obtain the scaling of the correlation length of the fluctuations,
$\lambda_c$: $\lambda_c^2=[\int\d y\ \rho_{2c}(x,y;t)]^{-1}\int\d y\ (x-y)^2\rho_{2c}(x,y;t)$. We obtain:
\beq
\lambda_c(t)\sim\kappa_H^{1/2}t^H.
\label{eq10}
\eeq
We thus see that the slow-down of the fluctuation build-up in Eq. (\ref{eq9}), with respect to 
a case without migration: $\rho_{2c}(x,x;t)=2n_0\Gamma t$ (Galton-Watson dynamics \cite{harris63}),
is precisely the smearing of the
fluctuations that is produced by anomalous diffusion. We notice that clustering is possible
(in one dimension) only as long as $H<1$; thus superdiffusive processes that are faster
than ballistic are excluded.

The derivation leading from Eq. 
(\ref{eq2}) to Eq. (\ref{eq9}) can easily be generalized to $D>1$, exploiting the fact
that, for Gaussian statistics, the PDF's entering Eq. (\ref{eq2}) are just product
of the PDF's of the individual components
of the position vectors $\x_1,\z$, etc. 
The two space integrals in Eq. (\ref{eq2}) eliminate the normalizations in 
$G(\z-\w)$ and in one of the $\rho_1(\x_{1,2},t|\z,\tau;\w,0)$. This leads to replace the 
$(\sigma(t;\tau))^{-1}$ factor in Eq. (\ref{eq7}) with $(\sigma(t;\tau))^{-D}$, 
and produces the generalization of Eq. (\ref{eq7}):
$\rho_2(\x,\x;t)\sim t^{1-DH}$.

\section{L\'evy flights}
In this case we do not have to bother with memory transfer among generations, 
as the process is Markovian. The process can be obtained as the continuous limit 
of a dynamics in which the individuals, at time intervals $\Delta t$, 
carry on independent jumps distributed with a PDF characterized by power-law tails: 
\beq
G(y,\Delta t)\sim|y|^{-1-\beta},
\qquad
0<\beta<2
\label{eq11}
\eeq
(L\'evy flights).
In Fourier space, this corresponds to the asymptotic behavior for small 
wavenumbers:
\beq
G_k(\Delta t)\simeq 1-c|k|^\beta\Delta t,
\label{eq12}
\eeq
where $c\sim(\Delta x)^\beta/\Delta t$, and $\Delta x$ is a microscale below which 
the scaling in Eq. (\ref{eq11}) ceases to hold.
The slow decay in $y$ of the PDF, Eq. (\ref{eq11}),  implies divergence of the second moment 
of the displacement
$\langle |y(t)|^2\rangle=\infty$ even for a single jump, and the
PDF $G(y,t)$ can be shown to tend to a L\'evy distribution; in 
Fourier space \cite{klafter87}:
\beq
G_k(t)=\exp(-c|k|^\beta t).
\label{eq13}
\eeq
Equation (\ref{eq13}) tells us that, 
although $G(y,t)$ does not have a second moment, we can identify 
a characteristic scale:
\beq 
y(t)=x(t)-x(0)\sim (ct)^{1/\beta},
\label{eq14}
\eeq
with the exponent $1/\beta$ playing a role analogous to that of the Hurst exponent for a 
diffusive process (more precisely, for a superdiffusive process, as $1/\beta>1/2$).

As in the previous section, let us assume that the individuals undergo processes of birth
and death with equal rate $\Gamma$, and that the individuals are distributed
at time $t=0$ with uniform density $n_0$. As in the previous case, it is easy to see
that the mean density remains constant $\rho_1(x,t)=n_0$, but fluctuations of increasing
amplitude, accounted for by the connected correlation $\rho_{2c}(x_1,x_2;t)$, are generated. 

The Markovian nature of the process allows us to derive a version of the evolution equation 
for $\rho_{2c}$, Eq. (\ref{eq2}), that is local in time:
\begin{widetext}
\beq
\rho_{2c}(x_1,x_2;t+\Delta t)&=&\int\d y_1\int\d y_2\ G(x_1-y_1,\Delta t)G(x_2-y_2,\Delta t)
\rho_{2c}(y_1,y_2;t)+2\Gamma n_0\Delta t\delta(x_1-x_2).
\label{eq15}
\eeq
\end{widetext}
Fourier transforming in $x_{1,2}$ and defining $\rho_{2c,k_1k_2}(t)=
2\pi\delta(k_1+k_2)C_{k_1}(t)$, we obtain
\beq
C_k(t+\Delta t)&=&G^2_{2k}(\Delta t)C_k(t)+2\Gamma n_0\Delta t.
\label{eq16}
\eeq
For $t\gg\Delta t$, the space scale contributing in Eq. (\ref{eq16}) are those corresponding
to the asymptotics in Eq. (\ref{eq11}). We can thus use Eq. (\ref{eq12}), and
Eq. (\ref{eq16}) becomes, taking the continuous limit:
\beq
\dot C_k+2c|k|^\beta C_k=2\Gamma n_0,
\label{eq17}
\eeq
that can be seen as a form of forced fractional heat equation
\cite{metzler00}.
Solution by Laplace transform in time gives us:
\beq
C_{kz}=\frac{2\Gamma n_0}{z(z+2c|k|^\beta)}.
\label{eq18}
\eeq
Inverse Fourier transform at zero space separation gives then the result
\beq
C_z(0)=
\int\frac{\d k}{2\pi}C_{kz}
=B\Gamma n_0c^{-1/\beta}z^{1/\beta-2},
\label{tmp}
\eeq
where $B=\int_0^{+\infty}(1+2h)^{-1}h^{1/\beta-1}\d h$, and this
corresponds to the long-time asymptotics
\beq
\rho_{2c}(x,x;t)\sim \Gamma n_0c^{-1/\beta} t^{1-1/\beta}.
\label{eq19}
\eeq
A realization 
of such a process is shown in Fig. \ref{fbmfig2}.
Again, fast processes for which $\beta<1$ do not lead to clustering. 

In spite of the fact that the dynamics considered in this and the previous section 
are completely different, as clear from Eqs. (\ref{eq9}) and (\ref{eq19}), the
fluctuation build-up occurs in the two cases in the same way.
Looking at Eq. (\ref{eq18}) and comparing with Eqs. (\ref{eq7}) and (\ref{eq10}), 
we find in fact the same mechanism of smearing of fluctuations at scale $y(t)\propto t^{1/\beta}$
[Eq. (\ref{eq14}), L\'evy flights] or $\lambda_c(t)\propto t^H$ [Eq. (\ref{eq10}), Gaussian
diffusion].

As in the case of Gaussian diffusion, 
these results can be generalized to $D>1$. Replacing $\d k\to k^{D-1}\d k$
in the integral of Eq. (\ref{tmp}), leads,  from Eq. (\ref{eq18}), to
$C_z(0)\sim z^{-D/\beta-2}$. 
Again we find the result: $\rho_2(\x,\x;t)\sim t^{1-D/\beta}$.

\section{Conclusion}
We have provided two separate examples of how the scaling in the build-up of 
demographic fluctuations in a birth-death model, can be made dependent on the kind
of diffusion utilized for migration. We have seen that the
basic ingredient for such dependence is the smearing of the fluctuations by the diffusion process, 
provided the inclusion of demography does not affect migration at population scale.
With this we intend the fact that the dispersion of a group
of individuals has the same scaling in time independently of whether demography
is taken into account or not. 

As in the case of Brownian bugs, working in higher dimension $D>1$, would decrease
the build-up exponent in proportion; namely, in place of Eq. (\ref{eq9}) or (\ref{eq19}),
we find:
$\rho_{2c}(x,x;t)\propto t^{1-DH}$ and $\rho_{2c}(x,x;t)\propto t^{1-D/\beta}$
in the two cases.
We thus see that superdiffusion would lead to clustering only for $D=1$.
However, in contrast with the case of Brownian bugs, power-law growth of the
fluctuations remains possible for
$D>1$, in the case of sufficiently slow subdiffusion (at least if diffusion 
is produced by a Gaussian process).

Summarizing, and including the results in \cite{olla12} as well, the various
dynamical mechanism leading to anomalous diffusion, and the different ways
in which memory is transferred between different generations of individuals, 
produce the following results:
\begin{itemize}
\item 
Gaussian anomalous diffusion (FBM, individuals moving with velocity that
is solution of a GLE, and others): if offsprings share memory of the trajectories
with their parents, demographic fluctuations 
will scale anomalously, as described by Eq. (\ref{eq9}). Otherwise, the case
of Brownian bugs is recovered. 
\item
L\'evy flights: demographic fluctuations scale anomalously, as described
by Eq. (\ref{eq19}). 
\item
CTRW: in the absence of memory transfer between generations, the case of
Brownian bugs is recovered. If the offsprings share their
escape time (the only thing that they can share) with their parents,
the dynamics falls back on that of a Galton-Watson process.
\item
A spatial assembly of random traps: the same behavior of the CTRW with memory
transfer between generation is obtained; the dynamics falls back on 
that of a Galton-Watson process.
\end{itemize}

\acknowledgements This research was funded in part by Regione Autonoma della Sardegna.



\begin{thebibliography}{10}






\bibitem{zhang90}
Y.-C. Zhang, M. Serva and M. Polikarpov,
J. Stat. Phys. {\bf 58}, 849 (1990).

\bibitem{meyer96}
M. Meyer, S. Havlin and A. Bunde,
Phys. Rev. E {\bf 54}, 5567 (1996).

\bibitem{houchmandzadeh08}
B. Houchmandzadeh,
Phys. Rev. Lett. {\bf 101}, 078103 (2008) 

\bibitem{young01}
W.R. Young, A.J. Roberts and G. Stuhne,
Nature {\bf 412}, 328 (2001).

\bibitem{heinsalu10}
E. Heinsalu, E. Hern\'andez-Garc\'\i a and C. L\'opez,
Europhys. Lett. {\bf 92}, 40011 (2010)

\bibitem{olla12}
P. Olla,
Phys. Rev. E {\bf 85}, 021125 (2012)  

\bibitem{klafter87}
J. Klafter, A. Blumen and M.F. Shlesinger,
Phys. Rev. A {\bf 35}, 3081 (1987).

\bibitem{montroll65}
E.W. Montroll and G.H. Weiss,
J. Math. Phys. {\bf 6}, 167 (1965).

\bibitem{harris63}
T. Harris, {\it The theory of branching processes} (Springer-Verlag, Berlin, 1963).

\bibitem{montroll79}
E.W. Montroll and B.J. West, 
{\it Fluctuation phenomena}, eds. E.W. Montroll and J.L. Lebowitz (North-Holland,
Amsterdam 1979)

\bibitem{bouchaud90}
J.P. Bouchaud and A. Georges,
Phys. Rep. {\bf 195}, 128 (1990)

\bibitem{zwanzig}
R. Zwanzig, 
{\it Nonequilibrium statistical mechanics} 
(Oxford University Press, New York, 2001)

\bibitem{mandelbrot68}
B.B. Mandelbrot and J.W. Van Ness, 
SIAM Rev. {\bf 10}, 422 (1968).

\bibitem{janssen99}
H.K. Janssen, K. Oerding,  F. van Wijland and  H.J. Hilhorst,
Eur. Phys. J. B {\bf 7}, 137 (1999)

\bibitem{clark99}
J.S. Clark, M. Silman, R. Kern, E. Macklin and J. Hille Ris Lambers,
Ecology {\bf 80}, 1475 (1999)

\bibitem{sims08}
D.W. Sims {\it et Al.},
Nature {\bf 451}, 1098 (2008).

\bibitem{levandowsky97}
M. Levandowsky, B.S. White and F.L. Schuster,
Acta Protozool. {\bf 36}, 237 (1997)

\bibitem{houch02}
B. Houchmandzadeh,
Phys. Rev. E {\bf 66}, 052902 (2002).

\bibitem{note}
Recall that $\rho_1(x,t_2|z,t_1;w,0)=\rho_{y(t_2),y(t_1)}(x-w,z-w)/G(z-w,t_1)$,
where $\rho_{y(t_2),y(t_1)}$ is the joint PDF of the individual displacement
at times $t_{1,2}$, that is a Gaussian.

\bibitem{nota0}
We can write in general:
$\sigma(t;\tau)=|t-\tau|^Hf(\tau/t)$ with $f(w)>0$ for $0\le w\le 1$. This gives us 
$\int_0^t(\sigma(t;\tau))^{-1}\d\tau=t^{1-H}\int_0^1[(1-w)^Hf(w)]^{-1}\d w$.

\bibitem{biferale97} L. Biferale, G. Boffetta, A. Celani, A. Crisanti and A. Vulpiani,
Phys. Rev. E {\bf 57}, R6261 (1998)

\bibitem{metzler00}
R. Metzler and J. Klafter,
Phys. Rep. {\bf 339}, 1 (2000)

\end{thebibliography}
\end{document}